\documentclass[aps,prb,amsfonts,floatfix,superscriptaddress, twocolumn, 10pt]{revtex4-2}
\usepackage[utf8]{inputenc}
\usepackage{amssymb,amsmath}
\usepackage{bbm}
\usepackage{bm}
\usepackage{multirow, array}
\usepackage{float}
\usepackage[colorlinks=true,allcolors=blue]{hyperref}
\usepackage{graphicx}
\usepackage[normalem]{ulem}
\usepackage{xcolor}
\usepackage{appendix}
\usepackage[autostyle]{csquotes}

\newcommand{\kt}[1]{\ensuremath{|#1\rangle}}
\newcommand{\br}[1]{\ensuremath{\langle #1|}}
\newcommand{\bk}[2]{\ensuremath{\langle #1|#2\rangle}}

\newcommand{\HS}{\mathcal{H}}

\newcommand{\refEq}[1]{Eq.~(\ref{#1})}
\newcommand{\sectionTitle}[1]{\textit{#1.---}}
\renewcommand{\vec}[1]{\mathbf{#1}}
\newcommand{\refApp}[1]{\cite{theelSupplementalMaterialTitle}}
\newcommand{\refFig}[1]{Fig.~\ref{#1}}
\newcommand{\refFigs}[1]{Figs.~\ref{#1}}
\newcommand{\refAppSec}[1]{Sec.~\ref{#1}}

\begin{document}

\title{Chirally-protected state manipulation by tuning one-dimensional statistics}

\author{F. Theel}
\affiliation{Center for Optical Quantum Technologies, Department of Physics, University of Hamburg, Luruper Chaussee 149, 22761 Hamburg Germany }
\author{M. Bonkhoff}
\affiliation{I. Institute for Theoretical Physics, Universit{\"a}t Hamburg, Notkestraße 9, 22607 Hamburg, Germany}
\author{P. Schmelcher}
\affiliation{Center for Optical Quantum Technologies, Department of Physics, University of Hamburg, Luruper Chaussee 149, 22761 Hamburg Germany }
\affiliation{The Hamburg Centre for Ultrafast Imaging, University of Hamburg, Luruper Chaussee 149, 22761 Hamburg, Germany}
\author{T.~Posske}
\affiliation{I. Institute for Theoretical Physics, Universit{\"a}t Hamburg, Notkestraße 9, 22607 Hamburg, Germany}
\affiliation{The Hamburg Centre for Ultrafast Imaging, University of Hamburg, Luruper Chaussee 149, 22761 Hamburg, Germany}
\author{N.L.~Harshman}
\affiliation{Physics Department, American University, Washington, DC 20016, USA}

\begin{abstract}
Chiral symmetry is broken by typical interactions in lattice models, but the statistical interactions embodied in the anyon-Hubbard model are an exception. This is an example for a correlated hopping model where chiral symmetry protects a degenerate zero-energy subspace. Complementary to the traditional approach of anyon braiding in real space, we adiabatically evolve the statistical parameter and find non-trivial Berry phases and holonomies in this chiral subspace. The corresponding states possess stationary checkerboard patterns in their $N$-particle densities which are preserved under adiabatic manipulation. We give an explicit protocol for how these chirally-protected zero-energy states can be prepared, observed, validated, and controlled.
\end{abstract}

\maketitle

In two dimensions, Abelian braid anyons with fractional exchange statistics arise from the topological analysis of configuration spaces~\cite{leinaasTheoryIdenticalParticles1977, goldinParticleStatisticsInduced1980, wilczekQuantumMechanicsFractionalSpin1982, wuGeneralTheoryQuantum1984, biedenharnAncestryAnyon1990, khareFractionalStatisticsQuantum2005}, and non-Abelian anyons of various forms have been proposed as the working material for robust topological quantum computing protocols~\cite{kitaevAnyonsExactlySolved2006, nayakNonAbelianAnyonsTopological2008}. 
However, non-standard exchange statistics are not an exclusively two-dimensional phenomenon.
Their key features have been proposed and investigated in one-dimensional systems since the beginning of the field~\cite{leinaasTheoryIdenticalParticles1977, haldaneFractionalStatisticsArbitrary1991, hanssonDimensionalReductionAnyon1992, agliettiAnyonsChiralSolitons1996, kunduExactSolutionDouble1999, phamAnyonsOneDimension2000, batchelorBetheAnsatz1D2007, keilmannStatisticallyInducedPhase2011, posskeSecondQuantizationLeinaasMyrheim2017, greiterAnyonsOneDimension2022, harshmanTopologicalExchangeStatistics2022, nagiesBraidStatisticsConstructing2024}, leading to experimental proposals~\cite{longhiAnyonsOnedimensionalLattices2012, yannouleasAnyonOpticsTimeofflight2019, schweizerFloquetApproachZ22019, gorgRealizationDensitydependentPeierls2019, zhangAnyonicBoundStates2023, bonkhoffCoherencePropertiesRepulsive2023} and recent realizations in Raman-coupled Bose-Einstein condensates~\cite{chisholmEncodingOnedimensionalTopological2022, frolianRealizing1DTopological2022}. This, for the first time, opens the possibility to tune the statistical angle of anyons in an experimentally accessible platform.

The anyon-Hubbard model offers the possibility for this exploration~\cite{keilmannStatisticallyInducedPhase2011, greschnerAnyonHubbardModel2015, straterFloquetRealizationSignatures2016, bonkhoffBosonicContinuumTheory2021}. Exchange statistics described by a statistical angle $\theta$ that interpolates between bosons $\theta=0$ and fermions $\theta = \pi$ are implemented on a one-dimensional lattice using Floquet-manipulated Rb atoms in quantum gas microscopes~\cite{Kwan2024} by density-dependent Peierls phases  ~\cite{greschnerDensityDependentSyntheticGauge2014,cardarelliEngineeringInteractionsAnyon2016,clarkObservationDensityDependentGauge2018}.
Such phases lead to intriguing effects, including statistically induced phase transitions \cite{keilmannStatisticallyInducedPhase2011, greschnerAnyonHubbardModel2015, langeAnyonicHaldaneInsulator2017}, quasi-condensation at finite momenta \cite{keilmannStatisticallyInducedPhase2011,greschnerAnyonHubbardModel2015}, emerging Friedel oscillations \cite{straterFloquetRealizationSignatures2016,langeStronglyRepulsiveAnyons2017, bonkhoffBosonicContinuumTheory2021}, as well as asymmetrical transport and expansion dynamics. \cite{liuAsymmetricParticleTransport2018,greschnerProbingExchangeStatistics2018}.

\begin{figure*}[t]
    \includegraphics[width=0.9 \linewidth]{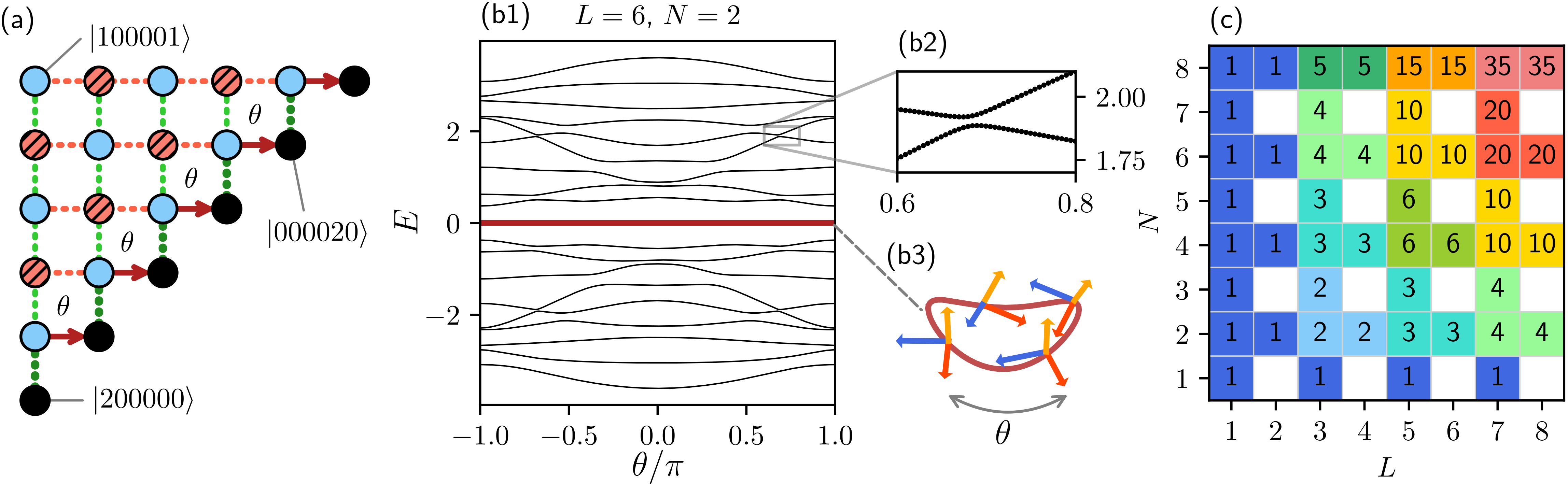}
	\caption{Configuration space representation of the anyon-lattice model and its chirally-protected zero-energy subspace $\HS_0$. 
    (a) Bosonic configuration space for $N=2$ particles on $L=6$ sites. Each dot represents a bosonic number state with doubly-occupied sites (black) and singly-occupied (not black) and with positive chirality (black, red hashed) or negative chirality (plain blue). The edges represent hopping processes with amplitude $-J$ (thin), $-\sqrt{2}{J}$ (thick, dotted), and $-\sqrt{2}Je^{i\theta}$ (arrow). Boundary plaquettes have a flux with a density-dependent Peierls phase $\theta$ through them.
    (b1) Energy spectrum as a function of the statistical angle $\theta$. The zero-energy subspace has dimension $d_0=3$ for $N=2$ and $L=6$ (red line) without avoided crossings (inset (b2)).   
     (b3) Schematic rotation of $\HS_0$ by cyclic manipulation of $\theta$.
    (c) The minimal degeneracy $d_0$ of $\HS_0$.
    }
 \label{fig:degeneracies}
\end{figure*}

Although the anyons realized by the anyon-Hubbard model are Abelian, their non-standard exchange statistics reveals topological structures in configuration space which we propose to exploit for non-Abelian state manipulation. To motivate this, consider that the density-dependent Peierls phases which implement the statistical interaction can induce synthetic magnetic fluxes through plaquettes in configuration space~\cite{Kwan2024, nagiesBraidStatisticsConstructing2024}. In the anyon-Hubbard model, these fluxes provide a phase $\exp(\pm i \theta)$ depending on the order in which the particles exchange; see \refFig{fig:degeneracies}(a). Many models in which underlying canonical particles experience correlated hopping processes share this feature, including~\cite{libertoQuantumSimulationCorrelatedhopping2014, greschnerDensityDependentSyntheticGauge2014, greschnerProbingExchangeStatistics2018,
cardarelliEngineeringInteractionsAnyon2016, hudomalQuantumScarsBosons2020, chuPhotonmediatedCorrelatedHopping2023, Serbyn2023}. Such correlated hopping processes can be engineered to break parity and time-reversal symmetry \cite{robnikFalseTimereversalViolation1986}.
However, correlated hopping models with nearest-neighbor hopping processes, such as the anyon-Hubbard model without on-site, Hubbard interactions, preserve the chiral symmetry associated with bipartite lattices~\cite{damskiMottinsulatorPhaseOnedimensional2006, grusdtTopologicalEdgeStates2013}. As a result, theorems about bipartite spatial lattices~\cite{sutherlandLocalizationElectronicWave1986, liebTwoTheoremsHubbard1989, ramachandranChiralFlatBands2017,Schomerus2020} can be generalized from real space to configuration space in models with chirally symmetric interactions and particle number conservation.

In this article, we show that one-dimensional lattice anyons with only statistical interactions host a degenerate zero-energy subspace protected by chiral symmetry. Adiabatically tuning the statistical angle $\theta$ from $0$ to $2\pi$ varies all energy levels in the spectrum except for the zero-energy modes, as shown in \refFig{fig:degeneracies}(b1-b2). In this space, the variation generates nontrivial holonomies~\cite{katoAdiabaticTheoremQuantum1950, wilczekAppearanceGaugeStructure1984}, i.e.,\ unitary transformations similar to braiding non-Abelian two-dimensional anyons around each other~\cite{nayakNonAbelianAnyonsTopological2008}. This scheme can be used for chirally-protected non-Abelian state preparation. Building on theorems of Lieb and Sutherland \cite{sutherlandLocalizationElectronicWave1986,liebTwoTheoremsHubbard1989}, we find that these results hold for the anyon-Hubbard model in the experimentally accessible limit of only statistical interactions~\cite{Kwan2024} and extend to chirally-symmetric correlated hopping models of bosons with particle number conservation. In these models, chiral symmetry equips the configuration space with a stationary checkerboard pattern that can be experimentally revealed by the $N$-body density correlations and the spatially off-diagonal correlation functions of zero-energy states and the quasi-momentum distribution. We furthermore show how these zero-energy states can be prepared from typical initial states and manipulated by steering the statistical angle.

\sectionTitle{Chiral symmetry and correlated hopping} For a particle-number conserving bosonic system, a chiral symmetry corresponds to a unitary operator $\hat{S}$ that anticommutes with the Hamiltonian $[\hat{S},\hat{H}]_+ = 0$~\cite{ludwigTopologicalPhasesClassification2015, yuSymmetryProtectedDynamical2017, xiaoRevealingSpatialNature2024}.  Such an operator is an involution $\hat{S}^2=1$ that partitions the finite-dimensional Hilbert space into chiral eigenspaces $\HS = \HS_+ \oplus \HS_-$ with eigenvalues $\chi = \pm 1$ and dimensions $\dim\HS_\pm = d_\pm$.
$\hat{H}$ is anti-block diagonal in the chiral basis and has a symmetric spectrum.
Therefore, the zero-energy eigenspace 
$\HS_0$ has dimension $d_0 \geq |d_+ - d_-|$~\cite{sutherlandLocalizationElectronicWave1986, liebTwoTheoremsHubbard1989, ramachandranChiralFlatBands2017, Schomerus2020, karleAreaLawEntangledEigenstates2021, nicolau2025}. This minimal degeneracy of the zero-energy subspace $d_0$ does not depend on the details of the Hamiltonian and is protected against perturbations that preserves chiral symmetry.

For $N$ spinless bosons on a one-dimensional lattice with $L$ sites, the operator $\hat{S}$ that realizes chiral symmetry for nearest-neighbor correlated-hopping models is~\cite{damskiMottinsulatorPhaseOnedimensional2006, grusdtTopologicalEdgeStates2013, yuSymmetryProtectedDynamical2017}:
\begin{equation}\label{eq:S}
  \hat{S} = \exp\left(i \pi \sum_{k=1}^L k \hat{n}_k \right),
\end{equation}
where $\hat{n}_k = \hat{b}^\dagger_k \hat{b}_k $ is the number operator on site $k$. The operator $\hat{S}$ transforms bosonic operators as $\hat{S} \hat{b}_k \hat{S} = (-1)^k \hat{b}_k$ and acts like a local gauge transformation in configuration space that assigns opposite chirality to adjacent number states, i.e., number states that differ by a single hop have opposite chirality. See \refFig{fig:degeneracies}(a) for a depiction of $L=6$ and $N=2$, where the chiral operator partitions the $21$-site configuration space lattice into a checkerboard of sublattices with $d_+\!=\!12$ and $d_-\! =\! 9$ chiral number states, respectively. Chiral symmetry together with particle-number conservation guarantees that this system has at least $d_0\!=\! |d_+ - d_-|= 3$ zero-energy states with positive chirality, although the specific subspace $\HS_0 \subset \HS_+$ spanned by these three states depends on the Hamiltonian. Note that additional pairwise zero-energy degeneracies can appear due to other non-Abelian symmetries or special accidental degeneracies. In particular, the non-interacting Bose-Hubbard model with $L=8$ and $N \geq 3$ has additional degeneracies at $E=0$ because of the trigonometric relation $\cos(\pi / 9) +  \cos(5\pi / 9) +  \cos(7\pi / 9) =0$. These accidental, cyclotomic degeneracies are akin to the Pythagorean degeneracies of the infinite square well. They occur in pairs that split upon variation from $\theta=0$ and are therefore not chirally-protected.

On the zero-energy subspace, the chiral operator $\hat{S}$ is promoted to an actual symmetry, because $\HS_0$ is a subspace of the majority chirality eigenspace $\HS_\pm$~\cite{sutherlandLocalizationElectronicWave1986, liebTwoTheoremsHubbard1989, Schomerus2020, karleAreaLawEntangledEigenstates2021, nicolau2025}. This implies that all correlation functions vanish which are not invariant with respect to $\hat{S}$. For example, the operator $\hat{b}^\dagger_{j} \hat{b}_{j+1}$ has odd chirality and therefore has a vanishing expectation value for every state in $\HS_0$ leading to the emergence of a checkerboard pattern that is stationary in time. Remarkably, we find for $N<L$ an analogous pattern in the $N$-particle densities.
States in the zero-energy subspace have support exclusively on the majority sublattice \cite{sutherlandLocalizationElectronicWave1986,liebTwoTheoremsHubbard1989,hatsugaiRevisitingFlatBands2021}, forming an higher-dimensional generalization of a checkerboard pattern. The scenario of spontaneous symmetry breaking in the thermodynamic limit is ruled out by Elitzur's theorem  \cite{elitzurImpossibilitySpontaneouslyBreaking1975,batistaGeneralizedElitzursTheorem2005,halperinHohenbergMerminWagner2019}.

For general $N$ and $L$, the dimension of the chirally-protected zero-energy subspace can be derived using combinatorics
\cite{Schomerus2020,Serbyn2021,Zhang2024} and is depicted in Fig.~\ref{fig:degeneracies}(c):
\begin{equation}\label{eq:d0}
    d_0 = \left\{ \begin{array}{ll}
    0 & \mbox{for $N$ odd and $L$ even}\\
   \frac{(\lceil L/2 \rceil + \lfloor N/2 \rfloor -1)!}{(\lceil L/2 \rceil -1)!\lfloor N/2 \rfloor!} & \mbox{else,}
    \end{array}\right.
\end{equation}
where $\lfloor A \rfloor$ and $\lceil A \rceil$ are the floor and ceiling function of $A$, respectively.
The majority chiral subspace is $\HS_+$ for $N$ even and $\HS_-$ for $N$ odd and $L$ odd.

\sectionTitle{Model}
The statistical interactions of the anyon-lattice model are chirally symmetric, in contrast to ordinary Hubbard-type interactions. They embody momentum-dependent  
interactions that are periodic in the statistical angle $\theta$. The model is defined in terms of anyonic operators $\hat{a}_j$ with deformed commutation relations that obey fractional exchange statistics
\begin{eqnarray}
    \label{eq:commrel}
    \hat{a}_j \hat{a}_k^\dag - e^{-i\theta \mathrm{sgn}(j-k)} \hat{a}^\dag_k \hat{a}_j &=& \delta_{jk},\nonumber\\
    \hat{a}_j \hat{a}_k - e^{i\theta \mathrm{sgn}(j-k)} \hat{a}_k \hat{a}_j &=& 0.
\end{eqnarray}
The Hamiltonian with $L$ sites takes the form 
\begin{equation}
\label{eq:Hamiltonian}
    H(\theta) =-J\sum_{j=1}^{L-1} \left( \hat{a}_{j+1}^\dag \hat{a}_j + \mathrm{h.c.}\right).
\end{equation}
The model is mapped to canonical bosons by a fractional Jordan-Wigner transformation $\hat{a}_j=\hat{b}_je^{i\theta\sum_{l<j}\hat{n}_l}$~\cite{keilmannStatisticallyInducedPhase2011}:
\begin{eqnarray}
\label{eq:Hkwan}
    \hat{H}(\theta) &=& -J \sum_{j=1}^{L-1} \left( \hat{b}_{j+1}^\dag e^{-i \theta \hat{n}_{j}}\hat{b}_j +\mathrm{h.c}\right).
\end{eqnarray}
This results in a correlated hopping process mediated by a density-dependent Peierls phase. The $U=0$ instantiation of the anyon-Hubbard model (\ref{eq:Hkwan}) has been successfully implemented in a Floquet-driven quantum gas microscope~\cite{Kwan2024}.

\sectionTitle{Properties of the zero-energy subspace} 
We investigate the properties of the zero-energy subspace by exact diagonalization and find the dimension of the zero-energy subspace $d_0$ typically assumes its minimal value  according to Eq.~(\ref{eq:d0}) as expected when the only symmetries are Abelian. Additional accidental degeneracies appear for special $N$ and $L$ at $\theta=0$ that are not chirally-protected and unsuitable for adiabatic manipulation.

We first construct a convenient basis for the zero-energy subspace at $\theta=0$ (non-interacting bosons) and use this as a experimentally-verifiable starting point for state manipulation. The single-particle operators
\begin{eqnarray}
\hat{c}_\nu &=& \sqrt{\frac{2}{L+1}}\sum_{k=1}^L \sin\left(q_{\nu} k\right) \hat{b}_k,\phantom{a}q_{\nu}=\frac{\pi \nu }{L+1}
   \label{eq:b_dagger_tilde}   
\end{eqnarray}
diagonalize the Hamiltonian $\hat{H} = \sum_{\nu=1}^L \epsilon_\nu \hat{c}^\dagger_\nu\hat{c}_\nu$ with 
$\epsilon_\nu = -2 J \cos\left(q_{\nu}\right)$. These operators satisfy the chirality relation $\hat{S}^{\dagger}\hat{c}_\nu\hat{S} = \hat{c}_{L-\nu+1}$ with $\epsilon_\nu = - \epsilon_{L - \nu + 1}$. 

From these $L$ single-particle states, we can construct a $d_0$-dimensional basis of $N$-particle non-interacting states with zero energy and definite chirality.  First, note that for $L$ odd the single-particle state $\kt{s} \equiv c^{\dagger}_{(L+1)/2}\kt{0}$ has zero energy and chirality $\chi_s = -1$. Second, two-particle states of the form 
\begin{equation}\label{eq:chiralpair}
    \kt{p(\mu)} \equiv \hat{S}^{\dagger}\hat{c}^\dag_\mu \hat{S} \hat{c}^\dag_\mu \kt{0}\ \mbox{for}\ \mu \in \lfloor L/2 \rfloor
\end{equation}
have zero energy and chirality $\chi_p = + 1$. 
There are precisely $d_0$ ways to distribute $N$ indistinguishable bosons among these chiral pair states $\kt{p(\mu)}$ and chiral single-particle states \kt{s}, and from these a standard basis for the zero-energy subspace can be built. As expected, these states exhibit a stationary checkerboard pattern in configuration space.

\begin{figure}[t]
    \centering
    \includegraphics[width=\linewidth]{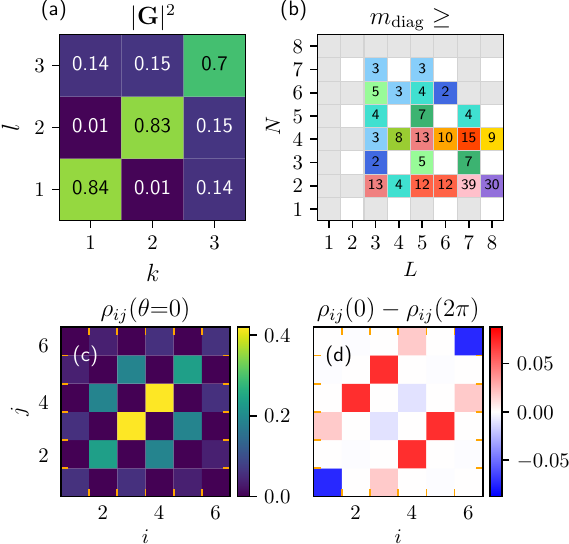}
	\caption{
    Holonomies for adiabatic evolutions from $\theta = 0$ to $2\pi$. 
    (a) The holonomy matrix $\mathbf{G}$ in pair basis $|p(\mu)\rangle$ (\refEq{eq:chiralpair}) reveals non-trivial adiabatic holonomies ( here $N=2$, $L=6$, and $N_\theta=10^4$ in \refEq{eq:holonomy_mat}, error $\sim \!10^{-3}$, see \refApp{}). 
    (b) The smallest integer $m_{\mathrm{diag}}$ such that $G^{m_\mathrm{diag}}$ is diagonal (within the error bounds) quantifies how nontrivial $\hat{G}$ is. $m_{\mathrm{diag}}$ increases strongly with system size.
    The two-body density of an initial zero-energy state (c), here $|p(1)\rangle$, alters significantly during the adiabatic evolution (d).
    \label{fig:adiab_proj}
    }
\end{figure}

\sectionTitle{Adiabatic manipulation of the zero-energy subspace} 
The periodic dependency in the statistical angle $\theta$ invites the question how the zero-energy subspace is affected under cyclic adiabatic variation of $\theta$. As the statistical angle is tuned, the zero-energy subspace $\HS_0$ moves and rotates within the larger subspace $\HS_\pm$ with majority chirality $\chi=\pm 1$. Similar to the Aharonov-Bohm effect on a ring, the zero-energy subspace accumulates topological Berry phases and nontrivial holonomies. These act as unitary transformations on the above-given standard basis vectors when tuning $\theta$ from $0$ to $2\pi$, as schematically depicted in \refFig{fig:degeneracies}(b3).

We propagate the zero-energy subspace from $\theta=0$ to $\theta = 2 \pi$ using Kato's adiabatic evolution~\cite{katoAdiabaticTheoremQuantum1950,Wilson1974}
\begin{equation}
    \hat{G} = \lim_{N_\theta \to\infty} \prod_{j=1}^{N_\theta} \hat{P}_j
    \label{eq:holonomy_mat},
\end{equation}
where $\hat{P}_j=\sum_{\mu=1}^{d_0} |\psi_\mu(2\pi j/N_\theta)\rangle \langle \psi_\mu(2\pi j/N_\theta)|$ is the projector onto the zero-energy subspace at the statistical angle $2\pi j/N_\theta$ and $|\psi_\mu\rangle$ is an orthonormal basis for the chiral zero-energy subspace; see \refApp{app:adiabatic_evolution_convergence} for numerical details.
The unitary matrix $\mathbf{G}_{\mu\mu'} = \br{\psi_{\mu'}} \hat{G} \kt{\psi_\mu}$ then embodies the adiabatic holonomy for a complete loop.

For $L=2$ and $N$ even where the zero-energy subspace is non-degenerate, we can exactly calculate the (Abelian) holonomy to $\mathbf{G} = \exp\left[i \pi N (N/2 +1)/4\right] = \pm 1$, see \refApp{app:Betheansatztwosites}.
More generally, we must investigate the non-Abelian holonomies of larger systems by exact diagonalization~\cite{theelSupplementalMaterialTitle,weisse2008, 2020SciPy-NMeth}.
As an example, in \refFig{fig:adiab_proj}(a), we present the holonomy matrix $\mathbf{G}$ in the chiral pair basis $\kt{p(\mu)}$ at $\theta = 0$ for a system with $N=2$ and $L=6$ and $d_0=3$.
The adiabatic evolution causes a significant non-trivial rotation of the zero-energy subspace during the adiabatic evolution, which we indicate by ${\rho}_{ij}^{\nu}\left(\theta\right)=\langle \nu(\theta) |\hat{b}_i^\dagger \hat{b}_j^\dagger \hat{b}_j \hat{b}_i | \nu (\theta)\rangle$, an observable accessible in the corresponding experiments  \cite{Kwan2024}. Additionally, note that an adiabatic scheme has been realized in a similar setup \cite{kimAdiabaticStatePreparation2024} and the experiment reported in \cite{Kwan2024} demonstrated a high control over the on-site interaction strength. 
We have checked that the holonomy matrix in \refFig{fig:adiab_proj}(a) is robust against small modulations of possible on-site interactions (see \refApp{}).
For example, we can take the initial two-body density $\rho_{ij}^{\nu}(\theta=0)$ of the basis state $\kt{p(1)}$ [\refFig{fig:adiab_proj}(c)] and compare with ${\rho}_{ij}^{\nu}(\theta = 2 \pi)$, i.e., the two-body density corresponding to the adiabatically propagated state $\hat{G}\kt{p(1)}$. This yields a significant difference in the densities, see \refFig{fig:adiab_proj}(d).
Interestingly, for variations of parameters other than $\theta$ that preserve chiral symmetry, such as local variations in hopping strength $J$, we find that the connection is flat, meaning that $\hat{G}$ is not altered. Consequently, the results for the holonomy matrix $\mathbf{G}$ are robust against fluctuations in the adiabatic manipulation.
Moreover, the zero-energy subspace gets larger as $N$ and $L$ increase, allowing for a richer structure of non-Abelian holonomies. However, the energy gap between the zero-energy states and the adjacent energy levels also narrows,
which increases the adiabatic time scale and enhances incoherence by diabatic noise.

Tuning the statistical angle therefore implements a topologically protected operation on the zero-energy subspace, a concept which has been extensively explored in the context of Majorana modes and other non-Abelian anyons \cite{nayakNonAbelianAnyonsTopological2008}. To quantify the nontriviality of $\mathbf{G}$, we determine the smallest integer $m_{\mathrm{diag}}$ such that $\mathbf{G}^{m_{\mathrm{diag}}}$ is diagonal, i.e., trivial in the context of state manipulation. Also, at least $m_{\mathrm{diag}}$ states in the computational space can be prepared from an initial state by repeated application of this operation \cite{kitaevUnpairedMajoranaFermions2001}. For instance, the braiding of Majorana modes becomes diagonal for $m_{\text{Maj.}} = 2$ and trivial with $G_{\text{Maj.}}^4 = 1$ \cite{nayakNonAbelianAnyonsTopological2008}. 
Within our error bounds, we find $m_{\text{diag}}>4$ for various particle numbers and system sizes, particularly for large ones, see \refFig{fig:adiab_proj}(b) and \refApp{}.

This idea can be elevated to non-Abelian holonomies if we implement the site-dependent statistical parameter $\theta_j$. This leads to a generalized anyonic exchange algebra  $a^\dagger_j a_k = \hat{a}_i\hat{a}_j^{\dagger}-e^{i\theta_{i,j}}\hat{a}_j^{\dagger} \hat{a}_i=\delta_{i,j}$ with  
$\theta_{i,j}= \theta_j$ for $i>j$, $\theta_{i,j}= -\theta_i$ for $i<j$, and $\theta_{i,i}= 0$.
For different exchange phases $\theta_L$ in the left and $\theta_R$ in the right part of the system \cite{theelSupplementalMaterialTitle, lauQuantumWalkTwo2022}, we generally find non-Abelian holonomies with $\left[\mathbf{G}_{L},\mathbf{G}_{R}\right]\neq 0$, meaning the fluctuation-protected state manipulation depends on which statistical angle is altered first \refApp{}.

\begin{figure}[t]
    \centering
    \includegraphics[width=0.9\linewidth]{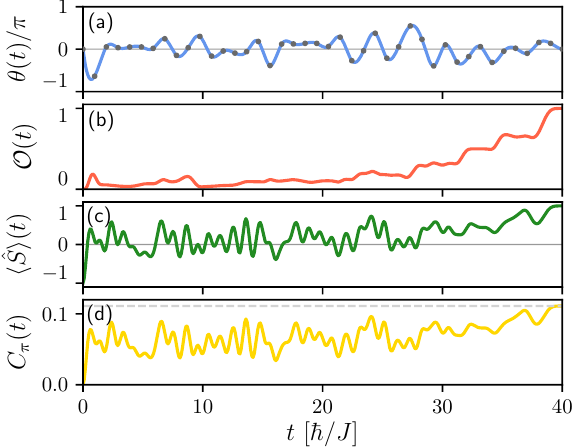}
	\caption{\label{fig:state_preparation}
    Steering to the zero-energy subspace, starting with $N=2$  bosons at the center of $L=6$ sites. 
    (a) The statistical phase $\theta(t)$ is dynamically varied along $60$ optimized interpolation points (gray) such that (b) the overlap $\mathcal{O}(t_f)$ of the time-evolved state with the zero-energy state $|p(1)\rangle$ (\refEq{eq:chiralpair}) is maximal, reaching a final overlap of $1-10^{-4}$. The time evolution of (c) the chirality and (d) the static structure factor $C_\pi(t)$, which approaches its maximal value $N^2/L^2$ (dashed line) indicating the emergence of a checkerboard pattern, the signature of the zero-energy subspace.
    }
\end{figure}

\sectionTitle{Steering to the zero-energy subspace}
The basis states for $\theta=0$ can be prepared by temporal variation of $\theta=\theta(t)$~\cite{Kwan2024}. As a simple, accessible initial state, we consider two bosonic particles at the central sites of a lattice with $L=6$ sites, i.e.,~$\kt{\Psi(t=0)}=\kt{001100}$.
In order to find a path for $\theta(t)$ which maximizes the overlap of the propagated wave function at time $t_f$ with a chiral pair target state,~$\mathcal{O}(t_f)=|\bk{p(\mu)}{\Psi(t_f)}|^2$, we represent $\theta(t)$ with $M$ interpolation points (for details see \refApp{app:numerical_details}). These $M$ points serve as an input for a gradient based optimization algorithm which updates the path $\theta(t)$ until $\mathcal{O}(t_f)$ becomes maximal \cite{byrdLimitedMemoryAlgorithm1995,zhuAlgorithm778LBFGSB1997,lewisOptimalControl2012}.

In \refFig{fig:state_preparation}, we present details of the optimization routine for $M\!=\!40$ and $t_f\!=\!40$. The chosen path for $\theta(t)$ shown in \refFig{fig:state_preparation}(a) reaches a fidelity $\mathcal{O}(t_f) =1-10^{-4}$ in \refFig{fig:state_preparation}(b), which can be further increased for larger $M$ and $t_f$. As a measure whether the final state converges to the zero-energy subspace, we monitor the expectation value of the chiral operator $\hat{S}$ in \refFig{fig:state_preparation}(c) and investigate the steering of the structure factor \cite{lewensteinUltracoldAtomsOptical2012} in \refFig{fig:state_preparation}(d)
\begin{align}
    C_{q}(t) = \frac{1}{L^2} \sum_{jk}e^{iq(j-k)}\langle \Psi(t) | \hat{n}_j \hat{n}_k |\Psi(t) \rangle.
    \label{eq:checkerboard}
\end{align}
We find a dominant signal at $q=0,\pi $ that reveals the presence of the checkerboard pattern in the $N$-particle density correlation for $L>N$, which is conveniently experimentally accessible in few-particle systems, see
\refFig{fig:adiab_proj}(c) \cite{kuhnerOnedimensionalBoseHubbardModel2000, rossonCharacterizingPhaseDiagram2020}.
If the checkerboard pattern remains constant for some time after the steering process, the propagated state is in the zero-energy subspace. For large particle numbers, we instead propose to probe the one-body density matrix $\langle \psi | \hat{b}_i^\dagger \hat{b}_j | \psi \rangle$ directly by the Fourier transform of the quasi-momentum distribution $\langle \tilde{b}^\dagger_k \tilde{b}_k \rangle$. The scheme presented in \refFig{fig:state_preparation} is robust against small deviations from the optimal path $\theta(t)$ and allows for small fluctuations of an on-site interaction strength (see \refApp{}).

\sectionTitle{Conclusions}
We have shown that chirally-symmetric number-preserving correlated hopping models with bosons accommodate at least $d_0$ degenerate zero-energy states. The dimension of this space of states is robust against any parameter variations that preserve chiral symmetry, although the space itself sweeps through the chiral majority subspace. For the anyon-lattice model, states in the zero-energy chiral subspace can be prepared from experimentally-accessible initial states and detected by the characteristic checkerboard pattern in configuration space with current quantum gas microscopy techniques~\cite{Kwan2024}. As the statistical angle is adiabatically tuned, the zero-energy space picks up a non-Abelian holonomy for each cycle. This paradigm of non-Abelian state preparation can find further applications with the space of chirally-symmetric correlated hopping models by tuning more than one cyclic parameter. As an example, we propose implementing this non-Abelian state preparation scheme within a spatially-inhomogeneous anyon-lattice model. More generally, we believe that Floquet-driven density-dependent Peierls phases offer a rich perspective for future exploration. Such models contain synthetic magnetic fluxes in few-body configuration space, and they provide an alternate path to understanding topological interactions in low dimensional systems.

The authors thank Andr\'e Eckardt and Peter Gr{\ae}ns Larsen for discussions on correlated hopping models and Joyce Kwan, Bryce Bakkani-Hassani, Perrin Segura, Yanfei li, and Annie Zhi for discussions of their anyon-Hubbard model experiment. T.P. thanks Ingo Runkel for comments on the local varying statistical angles.
M.B.\ and T.P.\ acknowledge
funding by the European Union (ERC, QUANTWIST, project number $101039098$). The views and opinions
expressed are however those of authors only and do
not necessarily reflect those of the European Union
or the European Research Council, Executive Agency.
T.P.\ and P.S.\ acknowledge support by the Cluster of Excellence “CUI: Advanced Imaging of Matter” of the Deutsche Forschungsgemeinschaft (DFG) – EXC
2056 – project ID $390715994$ and T.P.\ acknowledges funding of the DFG project
No.\ 420120155. 
Additionally, N.H.\ was supported by the Deutscher Akademischer Austauschdienst, the Centre for Ultrafast Imaging: Advanced Imaging of Matter of the University of Hamburg. 
\bibliographystyle{apsrev4-2}

\newpage

\pagebreak
\appendix
\onecolumngrid

\begin{center}
	\textbf{\large  Supplemental Material for ``Chirality-protected state manipulation by tuning one-dimensional statistics''}

	\vspace{3ex}

	\begin{minipage}{0.83\linewidth}
		In this Supplemental Material, we introduce the basic computational tools and discuss details of the key results discussed in the main text.
		In \refAppSec{app:numerical_details}, we discuss the numerical details of the diagonalization procedure. In Sections \ref{app:adiabatic_evolution_convergence} and \ref{app:potentiated_holonomy_matrix}, we comment on the convergence behavior of the adiabatic evolution and the error treatment relevant for potentiating the holonomy matrices. In \refAppSec{app:Betheansatztwosites}, we calculate the quantized  geometric phase of the exactly solvable two-site model and extend the adiabatic manipulation to locally varying statistical angles in \refAppSec{app:local}.
		In \refAppSec{app:numerical_details_steering}, we provide more details on the protocol used to steer an initial state into the zero-energy space.
	\end{minipage}

	\vspace{5ex}

\end{center}

\twocolumngrid

\setcounter{equation}{0}
\setcounter{figure}{0}
\setcounter{table}{0}
\setcounter{page}{1}
\makeatletter
\renewcommand{\theequation}{S\arabic{equation}}
\renewcommand{\thefigure}{S\arabic{figure}}
\renewcommand{\bibnumfmt}[1]{[S#1]}
\renewcommand{\citenumfont}[1]{S#1}

\section{Implementation of the Anyon-Hubbard model}
\label{app:numerical_details}

Here, we provide details about the numerical implementation of the Anyon-Hubbard model. The objective is to solve the Hamiltonian $\hat{H}$ in \refEq{eq:Hkwan} exactly via exact diagonalization \cite{weisse2008_supp}.
Therefore, we express $\hat{H}$ in terms of the number state basis ${|\vec{n}_i\rangle}_{i=1}^{\mathcal{N}}$, which results in the matrix $\mathbf{H}_{ij}= \langle \vec{n}_i | \hat{H} | \vec{n}_j \rangle$. The number states, $\vec{n}=(n_1, \dots, n_L)$, specify the occupation of $N$ particles distributed on $L$ lattice sites with $\sum_k n_k = N$. Assuming indistinguishable particles, there are $\mathcal{N}=(N+L-1)!/[N!(L-1)!]$ basis states which also defines the dimension of Hilbert space.
The largest system size we consider in this work has a Hilbert space dimension of $\mathcal{N}=6435$ which corresponds to the case where $L=8$ and $N=8$ and, thus, is well in the range of what is computationally feasible.

Apart from calculating the ground state properties, we also calculate the time evolution of an initial state for a time-dependent Hamiltonian as done within the steering process or for the adiabatic propagation.
The time evolution is conducted by integrating the time-dependent Schr\"{o}dinger equation, i.e., a set of coupled ordinary differential equations. We use an eighth-order Runge-Kutta method within the module \texttt{scipy.integrate.ode} with the \texttt{scipy} \cite{2020SciPy-NMeth_supp} version 1.10.1 and the input parameters $nsteps=10^8$, $atol=10^{-10}$ and $rtol=10^{-10}$.

\begin{figure}[t]
	\centering
	\includegraphics[width=0.9\linewidth]{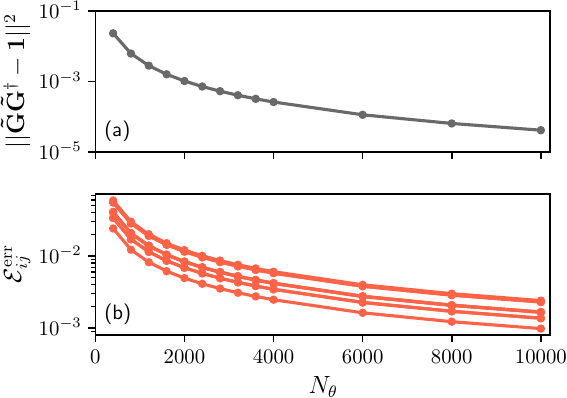}
	\caption{Convergence of the adiabatic holonomy when altering the statistical angle $\theta$ from $0$ to $2\pi$ ($L=6$ and $N=2$).
		(a) Distance of the holonomy matrix $\mathbf{\tilde{G}}$ to an unitary matrix.
		(b) Relative difference $\Delta \mathcal{E}_{ij}^{\mathrm{err}}$ between the entries of $\mathbf{\tilde{G}}$ obtained with $N_\theta$ projections and the unitary holonomy matrix, $\mathbf{G}$, obtained with $N_\theta=10^4$, see text for procedure.}
	\label{fig:adiab_convergence}
\end{figure}

\section{Convergence of the adiabatic evolution}
\label{app:adiabatic_evolution_convergence}

In the following, we determine the accuracy of the method used for obtaining the holonomy matrix $\mathbf{G}$ outlined in the main text for a system with $L=6$ lattice sites and $N=2$ particles corresponding to a three-fold degenerate zero-energy space. Within this method, the zero-energy states of $H(\theta=0)$ are adiabatically propagated by projecting the initial states consecutively onto zero-energy spaces lying on the path from $\theta=0$ to $2\pi$ \cite{katoAdiabaticTheoremQuantum1950_supp}.
In the adiabatic limit, $N_\theta \rightarrow \infty$, the overlap between the initial and adiabatically propagated states defines the holonomy matrix $\mathbf{G}$ (see main text). This matrix is unitary, i.e., $\mathbf{G}\mathbf{G}^\dagger=\mathbf{1}$ where $\mathbf{1}$ is the identity matrix.
This property can be probed for holonomy matrices $\mathbf{\tilde{G}}$ that are numerically obtained with a finite number of projection steps $N_\theta$ by measuring the distance $||\mathbf{\tilde{G}}\mathbf{\tilde{G}}^\dagger - \mathbf{1}||^2$. In \refFig{fig:adiab_convergence}(a), we show the convergence of this distance in dependence on the steps $N_\theta$ leading to a deviation from unitarity of around $\sim 10^{-4}$ for $N_\theta=10^4$.
As post process, we obtain the holonomy matrix $\mathbf{G}$ used for the analysis, by calculating the unitary matrix that is closest to the holonomy matrix $\mathbf{\tilde{G}}$ obtained with $N_\theta=10^4$ projections. For this, we first perform a singular value decomposition on $\mathbf{\tilde{G}}$, i.e., $\mathbf{\tilde{G}}=\mathbf{U}\mathbf{\Sigma} \mathbf{V}^\dagger$, where $\mathbf{U}$ and $\mathbf{V}$ are two unitary matrices and $\mathbf{\Sigma}$ is a diagonal matrix. Then we define the unitary holonomy matrix as $\mathbf{G}=\mathbf{U}\mathbf{V}^\dagger$, which has the property $||\mathbf{G} \mathbf{G}^\dagger - \mathbf{1}||^2=0$.

\begin{figure}[t]
	\centering
	\includegraphics[width=\linewidth]{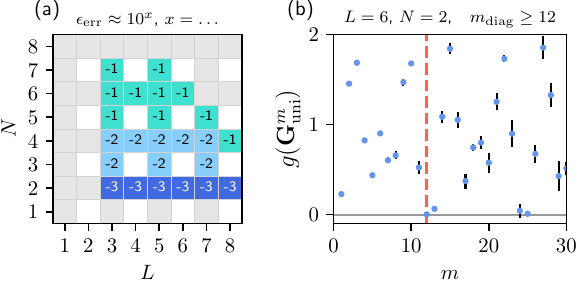}
	\caption{(a) Error of the holonomy matrix, $\epsilon_\mathrm{err}^\mathrm{proj}$, in dependence on the lattice sites $L$ and the particle number $N$. The error is defined as the largest deviation of the holonomy matrix obtained with $N_\theta=10^4$ projections to the closest unitary matrix (see \refAppSec{app:adiabatic_evolution_convergence}). (b) Calculating the distance of the potentiated holonomy matrix $\mathbf{G}^m$ to the diagonal matrix illustrated by $g(\mathbf{G}^m)$ [see \refEq{eq:dist_mat_id}]. We define $m_{\mathrm{diag}}$ as the smallest value of $m$ where $g(\mathbf{G}^m) - s_{g(\mathbf{G}^m)} <0$, i.e., for which the potentiated matrix cannot be distinguished from a diagonal matrix for a given error $s_{g(\mathbf{G}^m_\mathrm{proj-uni})}$ (red dashed line) (see \refAppSec{app:potentiated_holonomy_matrix}).}
	\label{fig:adiab_evolution_Gmat_m_err}
\end{figure}

On the basis of the unitary holonomy matrix, we analyze the convergence behavior of the single elements of $\mathbf{\tilde{G}}$ for different numbers of projections $N_\theta$. In particular, we show in \refFig{fig:adiab_convergence}(b) the relative difference $\mathcal{E}_{ij}^{\mathrm{err}}(N_\theta)= |\tilde{G}_{ij}^{N_\theta} - G_{ij}|/|G_{ij}|$, which reveals the expected convergence behavior for increasing $N_\theta$.
Based on this we quantify the error of the holonomy matrix as,
\begin{align}
	\epsilon_{\mathrm{err}} = \underset{ij}{\max} \left( \mathcal{E}_{ij}^{\mathrm{err}} \right),
	\label{eq:err_G}
\end{align}
i.e., as the largest deviation of the holonomy matrix obtained with $N_\theta=10^4$ projections, $\mathbf{\tilde{G}}$, from its closest unitary matrix, $\mathbf{G}$. In \refFig{fig:adiab_evolution_Gmat_m_err}(a) we show the order of the error in dependence of $L$ and $N$ (see also \refAppSec{app:potentiated_holonomy_matrix}).

Finally, we have compared the holonomy matrices obtained by applying projections as described to a method when the zero-energy states are propagated in time. More precisely, each zero-energy state in chiral pair basis representation is propagated in time while $\theta(t)$ is ramped linearly from $\theta(0)=0$ to $2\pi$ within a total propagation time of $t_f$. The accuracy of this method increases with the final propagation time $t_f$ and is exact in the adiabatic limit $t_f\rightarrow \infty$.
We have verified that the holonomy matrices obtained with both methods are in agreement.

\section{Potentiated holonomy matrices}
\label{app:potentiated_holonomy_matrix}
In \refFig{fig:adiab_proj}(b) of the main text, we present the minimal exponents $m_{\mathrm{diag}}$ of the holonomy matrix $\mathbf{G}$ in dependence of $L$ and $N$ for which the potentiated holonomy matrix, $\mathbf{G}^m$, cannot be distinguished from a unitary diagonal matrix.
We judge the distance of the potentiated holonomy matrix to the diagonal matrix by calculating
\begin{align}
	g(\mathbf{G}^m) = \sum_{ij} \left||G_{ij}^m|^2-\delta_{ij}\right|^2.
	\label{eq:dist_mat_id}
\end{align}
The value of $m_{\mathrm{diag}}$ is the smallest exponent $m$ for which $g(\mathbf{G}^m) <  s_m $, where $s_m$ denotes the numerical error. 
We estimate $s_m$ for a specific $L$-$N$ combination stochastically. To this end, we first calculate the relative absolute difference between the holonomy matrix and its closest unitary matrix, $\mathcal{E}_{ij}^{\mathrm{err}}$ (see \refAppSec{app:adiabatic_evolution_convergence}). Then we collect the values of $g$ for a set of purposely modified holonomy matrices, $(\mathcal{G}_k)_{ij}=G_{ij}^\mathrm{uni} + P_{ij}^k \Delta G_{ij}$, where $P_{ij}^k=\{-1, 1\}$. Thereby, $g$ is calculated for maximally $1000$ combinations of how to add/subtract the deviations $\mathcal{E}_{ij}^{\mathrm{err}}$ to/from $G_{ij}$. From the sample set $g(\mathcal{G}_k^m)$, we take the value with the largest absolute difference from $g(\mathbf{G}^m)$ as numerical error, i.e., $s_m=\underset{k}{\max}(|g(\mathcal{G}_k^m) - g(\mathbf{G}^m)|)$.
Evidently, the error increases with increasing particle numbers, which explains why the values for $m_{\mathrm{diag}}$ in these regimes are comparatively small, cf. \refFig{fig:adiab_proj}(b).
In \refFig{fig:adiab_evolution_Gmat_m_err}(b), we representatively show the evolution of $g(\mathbf{G}^m)$ for $L=6$ and $N=2$ and mark $m_{\mathrm{diag}}$ by a red dashed line.

\section{Exact Null State and Non-Degenerate Geometric Phase for $L=2$ and $N=2m$}
\label{app:Betheansatztwosites}

In this section, we derive exact results for a system with two sites, $L=2$, and an even number of particles, $N=2m$ with $m\in\mathbb{N}$, to gain intuition and to compare to the general, numeric results in the main text. 
In this case, the zero-energy space in \refEq{eq:d0} is one-dimensional.
In Ref.~\cite{bonkhoffCoherencePropertiesRepulsive2023_supp}, it was shown that the anyon-Hubbard dimer and the integrable Bose-Hubbard dimer are dual to each other, so in the following, we use this duality relation to calculate the geometrical phase for this special case analytically. For $U=0$ the zero-energy state of the Bose-Hubbard dimer in Fourier space reads
\begin{align}
	&\hat{H}=-2J(\hat{n}_0-\hat{n}_1),
	\label{eq.dimerzeroenergy}
	\\
	&\vert\Psi\rangle=\frac{1}{(N/2)!}(\hat{c}^{\dagger}_0)^{N/2}(\hat{c}^{\dagger}_1)^{N/2}\vert 0,0\rangle,
	\label{eq.dimerzeroenergystate}
\end{align}
with 
\begin{align}
	\hat{c}_n=\frac{1}{\sqrt{2}}\sum_{j=1}^{2}e^{i n j\pi }\hat{b}_j,\phantom{a} n=0,1.\label{eq.dimerzeroenergyfourier}
\end{align} 
By inversion of the Fourier transform and the usage of the duality transformation from Ref.~\cite{bonkhoffCoherencePropertiesRepulsive2023_supp}, we get the corresponding state of the anyon-Hubbard dimer in real space, with chirality $\chi=+1$
\begin{align}
	\vert\Psi\rangle=&\frac{1}{2^{N/2}((N/2)!)^2}\sum_{l=0}^{N/2}(-1)^{N/2-l}\sqrt{(N-2l)!(2l!)}\binom{N/2}{l}\nonumber
	\\
	&e^{i\theta\hat{n}_1(1+\hat{n}_2)}e^{i\theta(N+1))(\hat{n}_1-\hat{n}_2)/4}\vert N-2l,2l\rangle.
\end{align}
Thereby, the modes in \refEq{eq.dimerzeroenergyfourier} have an equal occupation of $N/2=m$, meaning that such states can exist only for even particle numbers in agreement with \refEq{eq:d0} in the main text.
As the zero-energy state is non-degenerate, the non-abelian holonomy reduces to a single phase, i.e.,
\begin{align}
	\phi=& \frac{1}{i\pi}\int_0^{2\pi}d\theta     \langle\Psi\vert\partial_{\theta}\vert\Psi\rangle
	=\frac{N/2\left(N/2+1\right)}{2},  \label{Zakphase}
\end{align}
that is indeed quantized as expected for chiral symmetric models \cite{hatsugaiQuantizedBerryPhases2006_supp, grusdtTopologicalEdgeStates2013_supp}. Interestingly, the value in \refEq{Zakphase} is exactly the proportionality factor of the Casimir operator of $SU(2)$, which is the dynamical symmetry group of the two-site problem.

\section{Locally varying $\theta$}
\label{app:local}

\begin{figure}[t]
	\centering
	\includegraphics[width=0.9\linewidth]{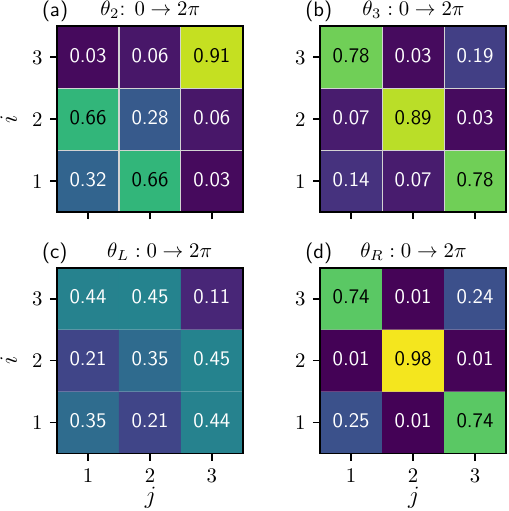}
	\caption{Absolute squares of the holonomy matrices $|\mathbf{G}_{\theta_j}|^2$ obtained after adiabatically steering a single or multiple $\theta_j$ from $0$ to $2\pi$ while the other statistical angles remain $0$ ($L=6$ sites and $N=2$). (a), (b) Tuning one statistical angle. (c), (d) Tuning the left side $\theta_L=\theta_1=\theta_2=\theta_3$ and the right side $\theta_R=\theta_4=\theta_5$ independently. The error of the shown holonomy matrices is everywhere of the order of $10^{-4}$.}
	\label{fig:adiab_evolution_local_theta}
\end{figure}

We next discuss a generalization of the adiabatic manipulation of statistical angles to multiple species of anyons and other chiral perturbations. To this end, we consider, the following Jordan-Wigner transformation with spatially varying $\theta_j$ \cite{lauQuantumWalkTwo2022_supp}, i.e.,
\begin{align}
	\hat{a}_j=\hat{b}_je^{i\sum_{l<j}\theta_l\hat{n}_l},
\end{align}
such that the following relation holds
\begin{align}
	\tilde{H}&=-J\sum_{j}\hat{a}^{\dagger}_j\hat{a}_{j+1}+\mathrm{h.c.}\nonumber
	\\
	&= -J\sum_{j}\hat{b}^{\dagger}_j\hat{b}_{j+1}e^{i\theta_j\hat{n}_j}+\mathrm{h.c.}.
\end{align}
Subsequently, we obtain \enquote{generalized} deformed commutation relations for the $\hat{a}_i$ particles, i.e.,
\begin{align}
	&\hat{a}_i\hat{a}_j^{\dagger}-e^{i\theta_{i,j}}\hat{a}_j^{\dagger}	\hat{a}_i=\delta_{i,j},\nonumber
	\\
	&\hat{a}_i\hat{a}_j-e^{-i\theta_{i,j}}\hat{a}_j\hat{a}_i=0\label{deformedalgebra},
	\\
	&\hat{a}_i^{\dagger}\hat{a}_j^{\dagger}-e^{-i\theta_{i,j}}\hat{a}_j^{\dagger}	\hat{a}_i^{\dagger}=0\nonumber,
\end{align}
with the statistical angle
\begin{equation}
	\theta_{i,j}=\begin{cases}      
		-\theta_i  & i<j, \\
		\theta_j   & i>j, \\
		0 &  \mathrm{else.}   \end{cases} \label{generalizedtheta}
\end{equation}
which generalize the deformed commutation relation for the particles in \refEq{eq:Hamiltonian} in the main text.

For illustration, we calculate the holonomy matrices $\mathbf{G}_{\theta_j}$  for a system with $L=6$ and $N=2$ when one or several statistical angles $\theta_j$ are varied from $0$ to $2\pi$. In \refFig{fig:adiab_evolution_local_theta}(a) and (b), we present the absolute squares of the holonomy matrix when either $\theta_2$ or $\theta_3$ is adiabatically tuned, respectively, while in (c) and (d) more than one $\theta_j$ is varied simultaneously, $\theta_1=\theta_2=\theta_3\equiv \theta_L$ and $\theta_4=\theta_5 \equiv \theta_R$, respectively. As shown, the holonomy matrices in \refFig{fig:adiab_evolution_local_theta} denote different rotations of the nullspace and are also to be distinguished from the case where all $\theta_j$ are tuned simultaneously, as shown in the main text in \refFig{fig:adiab_proj}(a). Moreover, we have checked that the shown holonomy matrices $\mathbf{G}_{\theta_j}$ are pair-wise non-commuting, $\mathbf{G}_{\theta_j} \mathbf{G}_{\theta_k} - \mathbf{G}_{\theta_k} \mathbf{G}_{\theta_j} \neq 0$ for $j\neq k$.

The adiabatic evolution has been done by applying a series of $N_\theta=10^4$ projections as discussed in \refAppSec{app:adiabatic_evolution_convergence}. We find for each shown holonomy matrix in \refFig{fig:adiab_evolution_local_theta} an error of $\epsilon_\mathrm{err}\approx 10^{-4}$, following the procedure of \refAppSec{app:adiabatic_evolution_convergence}.

\section{Steering process}
\label{app:numerical_details_steering}

In the following, we provide numerical details for the optimization routine employed for steering the system from a pure number state to a zero-energy eigenstate by optimally varying the statistical parameter $\theta(t)$ in time.
As mentioned in the main text, we represent $\theta(t)$ by a fixed number of $M_\mathrm{interp}+2$ time-wise equidistant interpolation points $\theta(t_i)$, where the first and last point remain fixed at $\theta=0$, i.e., $\theta(t_1)=\theta(t_{M_\mathrm{interp}+2})=0$. The interpolation employs a third degree spline function that passes through all interpolation points and is done with the module \texttt{scipy.interpolate.InterpolatedUnivariateSpline} from \texttt{scipy} \cite{2020SciPy-NMeth_supp} version 1.10.1.
In the next step, we choose an initial state $|\Psi(t=0)\rangle$ which serves as starting point for each time-propagation and an initial guess for the path $\theta(t)$. For the latter, we set all interpolation points, apart from the first and last, to $\theta(t_i)=0.01$, where $1 < i < M_\mathrm{interp}+2$, and calculate the respective interpolation.
The interpolation points are input to an optimization routine, where within each optimization step the corresponding time-propagation for $\theta(t)$ is calculated. After each propagation the overlap $\mathcal{O}(t_f)$ at the last time step between the propagated wave function with the target state is evaluated. The optimization routine varies the amplitudes of $\theta(t_i)$ after each propagation until the cost value $1-\mathcal{O}(t_f)$ becomes minimal. As optimization routine, we use the module \texttt{scipy.optimize.minimize} from \texttt{scipy} version 1.10.1 and the \textit{L-BFGS-B} method \cite{byrdLimitedMemoryAlgorithm1995_supp, zhuAlgorithm778LBFGSB1997_supp, lewisOptimalControl2012_supp}.

Note that altering the initial interpolation points can lead to different outcomes of the optimization routine regarding the final path of $\theta(t)$.
Moreover, we have checked that this optimization routine can be also applied to smaller system sizes than $L=6$, $N=2$.
Additionally, we considered different optimization routines such as a linear interpolation of $\theta$ (instead of a third degree spline) and a greedy optimization routine where the wave function is step-wise propagated according to the next optimal value for $\theta$ (instead of cyclically updating the optimal route for $\theta(t)$ after propagating the wave function from $t=0$ to $t_f$). However, both attempts were not successful since they both heave led to significant smaller overlaps with the target state than the procedure described above.
Another route we explored was to introduce onsite interactions $\frac{U(t)}{2} \sum_{i=1}^{L} \hat{n}_i(\hat{n}_i -1)$ to the Hamiltonian in \refEq{eq:Hkwan} in the main text and apply the optimization routine only to $U(t)$ and fix $\theta=0$.
However, also this method for the input parameters: $t_f=40$, $M_\mathrm{interp}=40$ and an initial $U(t_i)=0.1$ results in $\mathcal{O}(t_f)\approx 0.3$ for $L=5$ and $N=2$, while the procedure regarding an optimization with respect to $\theta(t)$ with the same input parameters leads to $\mathcal{O}(t_f)\approx 1 - 10^{-8}$.
Note that varying only the hopping parameter $J(t)$ while $U=\theta=0$ has no effect on the overlap which remains at $\mathcal{O}=0$.
We conclude that within our numerical studies varying $\theta$ dynamically remains the most promising method for zero-energy space state preparation.

Finally, we note that the zero-energy subspace can also be accessed using alternative cost functions. For example, we have verified that by varying $\theta$, it is possible to steer the system into the zero-energy subspace while requiring that the structure factor $C_\pi$ remains at its maximal value throughout the steering process, i.e., $C_\pi(t)\approx N^2/L^2$ and $\theta(t)=0$ for $t>t_f$. This observable remains stationary only within the chirally protected subspace. Moreover, any arbitrary off-diagonal correlation function can exhibit a similar behavior, depending on its chiral properties. Therefore, although the proposed measures may not uniquely identify the null space when considered individually, their combination leads to its identification.

\begin{figure}[h!]
	\centering
	\includegraphics[width=\linewidth]{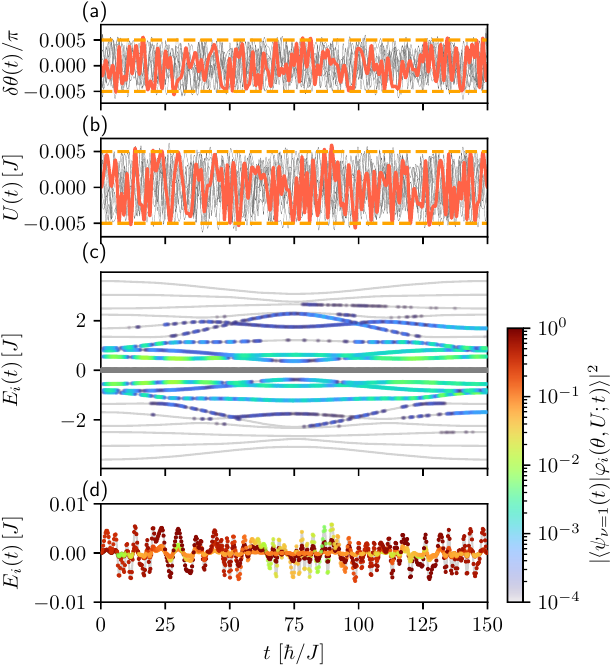}
	\caption{
		Here we exemplarily demonstrate that the adiabatic evolution can be appropriately approximated in terms of a time propagation which is robust with respect to additionally noise.
		We consider as initial state the zero-energy state $|p(\nu=1)\rangle$ and propagate the state for a varying statistical angle $\theta(t)=\theta_{\mathrm{lin}}(t) + \delta\theta(t)$ and on-site interaction strength $U(t)$ in time until $t_f=150\,[\hbar/J]$. The loop in control space is realized by $\theta_{\mathrm{lin}}(t)$, which is a linear function interpolating between $\theta_{\mathrm{lin}}(0)=0$ and $\theta_{\mathrm{lin}}(t_f)=2\pi$.
		We simulate small errors by adding small modulations $\delta\theta(t)$ to $\theta_{\mathrm{lin}}$, see red line in (a), and by considering finite modulations in $U(t)$, see red line in (b).
		The color-code in (c) and (d) quantifies the overlap of the propagated wave function $|\psi_{\nu=1}(t)\rangle$ with the $i$-th eigenstate $|\varphi_i(\theta, U; t)\rangle$ of the Hamiltonian $H(\theta, U;t)$ at time $t$. The gray lines in panels (c) and (d) denote the time-dependent eigenenergies $E_i$(t).
		Gray curves in (a) and (b) describe other paths of random errors used to calculate the black hashed tile in \refFig{fig:L6_N2_adiab_prop_err}. The distance between the orange horizontal dashed lines in (a) and (b) correspond to $\theta_{\mathrm{err}}$ and $U_{\mathrm{err}}$.
		A system with $L=6$ and $N=2$ is assumed corresponding to a three-fold degenerate zero-energy subspace, i.e., $d_0=3$. }
	\label{fig:L6_N2_prop_exc}
\end{figure}

\section{Studying the robustness of the adiabatic evolution and the steering protocol}
\label{app:robustness_U_theta}

In the following, we study the impact of finite random modulations of the statistical angle $\theta$ and the on-site interaction strength $U$ on the performance of the adiabatic evolution and the steering protocol. We account for the errors by considering time-dependent error functions $\delta\theta(t)$ and $U(t)$ when performing the time propagation.
Specifically, we model the error of the two-body on-site interactions by adding to the Hamiltonian, shown in \refEq{eq:Hkwan} of the manuscript, the interaction term $U(t)\sum_{i=1}^{L} \hat{n}_i(\hat{n}-1)/2$. The path of the time-dependent interaction strength, $U(t)$, is set by interpolating between random equidistant points $U_i=U(t_i)$ separated by $t_{i+1}-t_i=1$ and with the condition $U(0)=U(t_f)=0$. The points $U_i$ are chosen such that they lie within the interaction window $[-U_{\mathrm{err}}/2, U_{\mathrm{err}}/2]$, see for instance the orange dashed lines limiting the gray curves in \refFig{fig:L6_N2_prop_exc}(b) for $U_{\mathrm{err}}=0.01$. Analogously, we model the error for the statistical angle $\delta \theta(t)$ constrained by $\theta_{\mathrm{err}}$.

\subsection{Adiabatic evolution}

In order to demonstrate that the adiabatic evolution is indeed robust against small modulations of $\theta$ and $U$,
we perform the adiabatic evolution by propagating the zero-energy eigenvectors $|p(\nu)\rangle$ at $\theta=0$ in time for a varying statistical angle $\theta(t)=\theta_{\mathrm{lin}}(t) + \delta\theta(t)$ and on-site interaction strength $U(t)$.
The loop in control space is realized by $\theta_{\mathrm{lin}}(t)$, which is a linear function interpolating between $\theta_{\mathrm{lin}}(0)=0$ and $\theta_{\mathrm{lin}}(t_f)=2\pi$.

In Fig. \ref{fig:L6_N2_prop_exc}, we perform the time-propagation for a system consisting of $L=6$ lattice sites and $N=2$ particles prepared in the zero-energy eigenstate $|p(1)\rangle$ with error functions corresponding to strengths $\theta_{\mathrm{err}}=0.01\pi$ and $U_{\mathrm{err}}=0.1J$. A set of ten random error functions is depicted in \refFigs{fig:L6_N2_prop_exc}(a) and (b). In \refFigs{fig:L6_N2_prop_exc}(c) and (d) we plot as gray lines the eigenenergies of the system at a time $t$ depending on $\theta(t)$ and $U(t)$. The color-code denotes the probability to find the propagated wave function in one of the eigenstates. For visualization purposes, we mark the spectrum around $E=0$ in \refFig{fig:L6_N2_prop_exc}(c) with gray dots and zoom into this area with the appropriate color-code in \refFig{fig:L6_N2_prop_exc}(d).

Performing the adiabatic propagation with the error functions, we first notice that a finite $U$ breaks the degeneracy of the zero-energy subspace, as expected. Moreover, we find that the errors as well as the finite propagation time $t_f$ lead to small excitations of the neighboring spectrum around $E=0$.
To judge the impact of the applied error on the holonomy matrix, we perform the respective time-propagation for each zero-energy eigenstate $|p(\nu, t=0)\rangle$ and calculate the overlaps with the propagated states $|p(\nu, t=t_f)\rangle$, i.e., $\tilde{\mathbf{G}}'_{\mu,\nu}=\langle p(\mu, t_f) |p(\nu, 0)\rangle$. Subsequently, we obtain the closest unitary matrix to $\tilde{\mathbf{G}}'$ following the procedure outlined in \refAppSec{app:adiabatic_evolution_convergence} and yield $\mathbf{G}'$.
If we compare the holonomy matrix $\mathbf{G}'$ with the one obtained with the error-free method outlined in the main manuscript ($\mathbf{G}$), we find,
\begin{align}
	|\mathbf{G} - \mathbf{G}'|^2&= \begin{pmatrix}
		0.0006 & 0.0003 & 0.0001\\
		0.0003 & 0.0001 & 0.0001\\
		0.0001 & 0.0001 & 0.0000
	\end{pmatrix}.
\end{align}
Despite introducing substantial errors in $U$ and $\theta$, the deviations in the holonomy matrix elements turns out to be small. For larger propagation times and smaller errors, this agreement can be improved (see below).

\begin{figure}
	\centering
	\includegraphics[width=\linewidth]{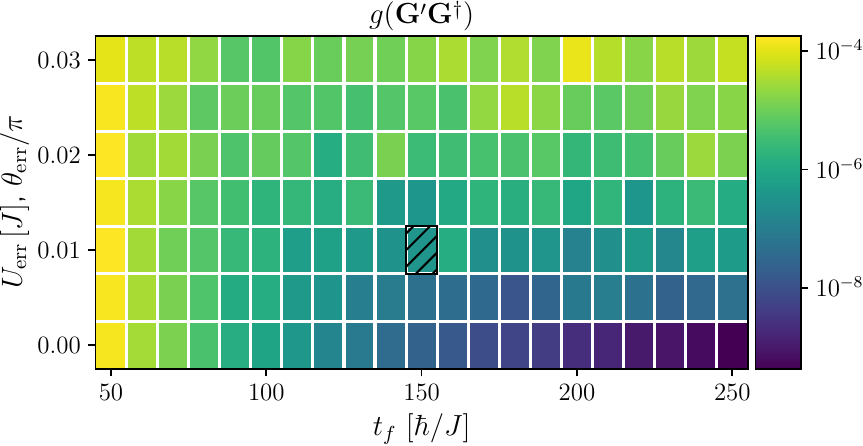}
	\caption{
		Averaged absolute-square differences $\overline{g}$ between the product $\mathbf{\mathbf{G}'G^\dagger}$ and the unit matrix, calculated using \refEq{eq:dist_mat_id}, dependent on the total propagation time $t_f$ and the strengths of the random errors $\theta_{\mathrm{err}}$ and $U_{\mathrm{err}}$. The behavior of $\overline{g}$ reveals small deviations between the holonomy matrices indicating a certain robustness of the applied adiabatic propagation method.
		The holonomy matrix $\mathbf{G}$ is obtained from the procedure introduced in the main text and \refAppSec{app:adiabatic_evolution_convergence}, while $\mathbf{G}'$ is obtained from an adiabatic time-propagation whose accuracy depends on $t_f$ and the artificially added errors $\delta \theta(t)$ and $U(t)$. The error functions are obtained by interpolating between random points which lie in the intervals $[-\theta_{\mathrm{err}}/2, \theta_{\mathrm{err}}/2]$ and $[-U_{\mathrm{err}}/2, U_{\mathrm{err}}/2]$, respectively. We average over results obtained for ten different random paths of $\delta \theta(t)$ and $U(t)$, see for instance the red and gray lines in \refFigs{fig:L6_N2_prop_exc}(a) and (b), which correspond to parameters marked by the black hashed tile.
		A system with $L=6$ and $N=2$ is assumed corresponding to a three-fold degenerate zero-energy subspace, i.e., $d_0=3$. }
	\label{fig:L6_N2_adiab_prop_err}
\end{figure}

We next work out the impact of the total propagation time.
Therefore, we quantify the deviation of the holonomy matrix $\mathbf{G}'$ from the reference matrix $\mathbf{G}$ by calculating the distance between the product $|\mathbf{G}'\mathbf{G}^\dagger|^2$ and the unit matrix using \refEq{eq:dist_mat_id}. In case the two matrices $\mathbf{G}'$ and $\mathbf{G}$ are identical, their product corresponds to the unit matrix up to phase factor.
In \refFig{fig:L6_N2_adiab_prop_err}, we show the quantity $g(\mathbf{G}'\mathbf{G}^\dagger)$ [\refEq{eq:dist_mat_id}] averaged over ten random paths of $U(t)$ and $\delta\theta(t)$ in dependence of the final propagation times $t_f$ and the interaction strength windows defined by $U_{\mathrm{err}}$ and $\theta_{\mathrm{err}}$. In the absence of any additional errors, $\delta\theta(t)=U(t)=0$, the agreement with the reference holonomy matrix $\mathbf{G}$ is improved when increasing the total propagation time $t_f$, which is expected since $t_f \rightarrow \infty$ denotes the adiabatic limit. On the other hand, we observe the expected decrease of the accuracy of the time-evolved adiabatic propagation when considering a finite error. Here, it is important to note that the accuracy does not decreases abruptly, but rather declines in a systematic manner in this way revealing a certain robustness of the method. Moreover, we find that for a finite random error increasing $t_f$ does not necessarily lead to an improvement of the adiabatic process. Since with increasing propagation times the applied error triggers additional excitation processes, one has to find the optimal value for $t_f$ to yield the best agreement with the reference holonomy matrix $\mathbf{G}$.

\begin{figure}
	\centering
	\includegraphics[width=0.9\linewidth]{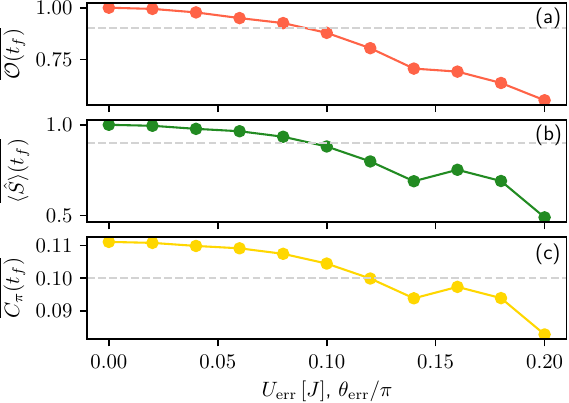}
	\caption{\label{fig:state_prep_vary_err}
		Demonstrating the robustness of the steering protocol against variations of the statistical angle $\theta$ and the on-site interaction strength $U$ in terms of (a) the overlap $\mathcal{O}(t)$ between the propagated wave function and the target state, (b) the chirality $\langle \hat{S}\rangle(t)$, and (c) the structure factor $C_\pi(t)$. Each dot represents an average over ten executions of the steering protocol evaluated at $t=t_f$. In each simulation a random error of $\delta\theta(t)$ and $U(t)$ is considered, where $\delta\theta(t)$ is added to the optimal path $\theta(t)$ [cf. \refFig{fig:state_preparation}(a)]. The random errors $\delta\theta(t)$ and $U(t)$ are interpolations of time-wise equidistant random points which lie within the interval $[-\theta_{\mathrm{err}}/2, \theta_{\mathrm{err}}/2]$ and $[-U_{\mathrm{err}}/2, U_{\mathrm{err}}/2]$.
		The horizontal gray dashed lines serves as guide for the eye and correspond to 90\% of the maximum value that the respective averaged observable can achieve.
		A system with $L=6$ and $N=2$ is assumed corresponding to a three-fold degenerate zero-energy subspace, i.e., $d_0=3$.}
\end{figure}

\subsection{Steering protocol}

Here, we comment on the robustness of the steering protocol with respect to deviations from the optimal path of $\theta(t)$ and for finite on-site interactions $U(t)$. In \refFig{fig:state_preparation}(a), we present the optimal path for $\theta(t)$ which, when applied, steers the initial number state $|001100\rangle$ into the zero-energy eigenstate $|p(1)\rangle$ for a system with $N=2$ particles distributed on $L=6$ lattice sites (see main text and \refAppSec{app:numerical_details_steering}).
To this function $\theta(t)$ we add a random error $\delta\theta(t)$ of the order of $\theta_{\mathrm{err}}$ and additionally, consider finite modulations of the on-site interaction $U(t)$ of the order of $U_{\mathrm{err}}$. After each execution of the steering protocol, we evaluate the overlap $\mathcal{O}(t_f)$ between the propagated wave function and the target state, the chirality $\langle \hat{S}\rangle(t_f)$, and the structure factor $C_\pi(t_f)$ at the time instance $t=t_f$. For each value of $\theta_{\mathrm{err}}/\pi = U_{\mathrm{err}}/J$ we average the expectation values over ten simulations and present the results in \refFig{fig:state_prep_vary_err}. As it can be seen, we only observe a small systematic decline of all observables once the error is increased reflecting the robustness of the applied steering protocol against small deviations of up to 5\%. Finally, as expected, the steering protocol breaks down for sufficiently large errors.

\bibliographystyle{apsrev4-2}

\end{document}